# The Observer in the Quantum Experiment

Bruce Rosenblum[1] and Fred Kuttner[1]

*A goal of most interpretations of quantum mechanics is to avoid the apparent intrusion of the observer into the measurement process. Such intrusion is usually seen to arise because observation somehow selects a single actuality from among the many possibilities represented by the wavefunction. The issue is typically treated in terms of the mathematical formulation of the quantum theory. We attempt to address a different manifestation of the quantum measurement problem in a theory-neutral manner. With a version of the two-slit experiment, we demonstrate that an enigma arises directly from the results of experiments. Assuming that no observable physical phenomena exist beyond those predicted by the theory, we argue that no interpretation of the quantum theory can avoid a measurement problem involving the observer.*

## 1. INTRODUCTION

From its inception the quantum theory had a "measurement problem" with the troubling intrusion of the observer. As usually seen, the problem is that the linearity of the Schrödinger equation forbids any system that it is able to describe from producing the unique result observed in an experiment. There is no mechanism in the theory--beyond the *ad hoc* probability assumption--by which the multiple possibilities given by the Schrödinger equation become the single observed actuality. The literature addressing this enigma has continued for decades and expands today. In fact, a list of the ten most interesting questions to be posed to a physicist of the future includes two questions directed to this enigma, one of which directly involves the observer[1].

Recent discussions have argued that with proper interpretation of the theory consideration of the observer is not needed to account for quantum phenomena[2]. What is being proposed can be unclear. Is the troublesome intrusion of the observer resolved merely for all *practical* purposes, or is it supposedly resolved in principle? Many readers take away the latter conclusion, which seems indeed to be implied.

Another recent opinion holds that the theory needs no interpretation at all[3]. Is it argued that the issue of the observer in quantum mechanics should be ignored? This is, again, the point readers may well come away with, that the issue of the observer is resolved or is inconsequential. That the issue is inconsequential is surely well taken for the practical use of the theory as a guide to the phenomena around us--which, we can assume, is the authors' actual point. But such a position is valid *only* when one wishes to put aside the philosophical implications suggested by the quantum theory, a perspective denying the profoundly influential worldview given us by classical physics. Interpreting quantum mechanics is a valid and potentially consequential enterprise.

Treatments of the measurement problem usually involve the mathematical structure of quantum theory. In contrast, we present a largely theory-neutral argument illustrating that the observer enters physics in the *experimental* situation logically prior

---

[1] Department of Physics, University of California, Santa Cruz, California 95064

to the quantum *theory*.  If we assume that no observable physical phenomena exist other than those specified by the present quantum theory, a role for the observer in the experiment can be denied only at the expense of challenging the belief that the observer makes free choices.  Therefore no interpretation of the present theory can establish a lack of dependence on the observer to the extent possible in classical physics.

The role of the observer was actually a problem in classical physics.  Given the determinism of Newtonian physics, the almost universal assumption of free will was early on seen as paradoxical.  With classical physics it was, however, a benign paradox.  The conscious mind receives information from the physical world only through eyes or other organs that are presumably understandable deterministically.  Conscious free will is manifest through deterministically understandable muscles.  The mind of the observer, that entity making free choices, being on the far side of eyes and muscles, could be considered an aspect of the universe isolated from the physical world to be treated by physics.  Since within that realm the different experiments which could be freely chosen by the observer never led to inconsistent pictures of the prior physical reality, classical physics could deal with only one part of a divided universe without considering the observer.

An analogous argument is not available for quantum physics.  Different quantum experiments that could be freely chosen by the observer *do* lead to inconsistent pictures of the prior physical reality.  This apparent intrusion of the free choice of the observer into the aspect of the physical world addressed by physics constitutes a measurement problem in the quantum experiment.  Our discussion will focus on this issue of the observer's choice.

Stapp emphasizes that a quantum measurement involves two choices[4].  The first is the choice by the observer of what experiment to do, that is, the choice of what question to ask of Nature. (Within the theory this involves the choice of basis.)  The second choice Stapp identifies is that by Nature giving the probabilistic answer to the experimenter's question, that is, providing a particular experimental outcome.  For reasons dating back to the 1927 Solvay Conference, Stapp calls the choice by the observer the "Heisenberg choice" and that by Nature the "Dirac choice," and we adopt this terminology.  Taking the example of the two-slit experiment, the Heisenberg choice might be the decision by the experimenter to find out either through which slit each particle comes, or in which maxima of the interference pattern each lands.  The Dirac choice by Nature would determine, in the first case, the particular slit, and for the second case, the particular maximum for each particle.

Most treatments of the measurement problem address the conflict between quantum theory's deterministic presentation of a superposition of macroscopically distinct states and the single actuality seen by the observer.  They thus focus on the Dirac choice.  Our focus on the Heisenberg choice, by contrast, exhibits the measurement problem arising directly from the experimental observations, logically prior to the theoretical concern.  Later on we will discuss how this latter problem can be transferred to the former.

## 2. A PARABLE

In discussions with physics colleagues, we find it hard to avoid the intrusion of the quantum theory and its interpretations into our attempt at a theory-neutral discussion of an experimental situation. Therefore we first describe a *fictional* classical experiment--something that really never happens. It's an analogy for the actually possible quantum experiment which will follow. The fully analogous aspect--that which is the same in both cases--is the reasoning leading to the experimenter's bafflement after viewing the experimental results. So here's our parable.

The Demonstrator displays a large number of pairs of boxes. She instructs the Experimenter to determine which box of each pair holds a marble by opening first one box of the pair and then the other. About half the time he finds a marble in the first box of the pair he looks into, and, if so, he finds the other box of that pair empty. Should the first box be empty, the other box of that pair always contains a marble. The Experimenter concludes that for this set of box pairs, one box of each pair contains a marble, and the other is empty; each marble is wholly within a single box.

The Demonstrator now notes that each marble can come apart into a white hemisphere and a black. Presenting a second set of box pairs, she instructs the Experimenter to determine for this set which box of each pair contains the white hemisphere and which the black by opening both boxes of each pair *at the same time.* The Experimenter always finds a white hemisphere in one of the boxes and a black in the other box of that pair. The Experimenter concludes that for this second set of box pairs, a marble is *distributed* over both boxes of each pair.

The Demonstrator now presents the Experimenter with further sets of box pairs and suggests that for each set he choose *either* of the two previous experiments. Allowed to repeat the experiment *of his choice* as many times as he wishes, the Experimenter always observes a result linked to the type of experiment he freely chooses: opening the boxes sequentially, he finds a whole marble in one box of a pair; opening boxes simultaneously, he finds a marble distributed over both boxes of the pair.

The puzzled Experimenter challenges the Demonstrator: "What if I had made the other choice? Before I opened the boxes each marble had to be either wholly in a single box or else have parts in both boxes." Her only reply, "That's an understandable assumption," has a condescending tone.

The Experimenter, sure there's some trickery, brings in a broad-based team of scientists and magicians (illusionists). However, after investigations which he accepts as exhaustive, they find no physical explanation. A psychologist on the team suggests the Experimenter may somehow have been led to choose the experiment corresponding to the marble's situation in the particular set of boxes. The Experimenter dismisses this; he *knows* his choices were freely made, that he could have made the other choice. A theorist on the team suggests: "There likely exists a field which is affected by how you open the boxes, and this field--though completely undetectable--creates the condition of the marble. We should seek a mathematical form for this field."

The Experimenter rejects this suggestion. "Even if one can write a mathematical form for an *undetectable* field, it's an untestable postulate. A 'completely undetectable' entity is not a scientific hypothesis."

At this point the psychologist offers an analysis: "I note a state of tension we call a 'cognitive dissonance.' It is brought about by your experiments displaying an inconsistency between two strongly-held beliefs: first, your belief that, prior to your observing it, the marble exists in a particular physical situation and, second, your belief that you exercised free choice. And, incidently, our theory of cognitive dissonance[2] [(5)] predicts attempts to resolve tension by adding a third belief. That 'completely undetectable field' seems such an addition."

The physicists present seem unimpressed by the psychological analysis.

## 3. THE EXPERIMENT

With a format similar to our parable, we describe a version of the two-slit experiment. We attempt a theory-neutral presentation of empirical observations. Of course no description of what happens--in a physics experiment, or in a courtroom--can be *completely* theory neutral, and we comment later on specifics of this lack of complete theory neutrality. We will confine our description to what is actually *seen*, what could be reported by any observer. Though the quantum theory is the standard explanation for the phenomena in this "quantum experiment," we describe the experimenter's actual observations without reference to quantum theory. By so doing, we characterize the measurement problem that arises directly from the experimental results.

However, for most readers, the observations described inevitably bring to mind the quantum theory. We wish to address these issues as soon as they likely arise, but to address them without departing from our theory-neutral presentation. The technique we use is to paragraph our quantum comments in bracketed italics and urge readers to consider them as *asides* to the theory-neutral story-line.

An Experimenter, whom we think of as a highly competent, open-minded observer unprejudiced by a particular theoretical doctrine, is presented with a set of box pairs, each pair containing a small object. The object could be any sufficiently small thing. Today it could be a photon, neutron, atom, or molecule. In the future, the objects may be considerably larger. To be general, we just refer to an object.

*[How the box pairs containing the object are prepared is irrelevant to our argument. But we can be explicit about a particular preparation method. Objects (wavepackets) are sent in from the left one at a time. They encounter a half-transmitting/half-reflecting mirror that divides each wavepacket into a transmitted horizontally traveling component and a reflected vertically*

---

[2] E. Aronson[(5)]: "Basically, cognitive dissonance is a state of tension that occurs whenever an individual simultaneously holds two cognitions (ideas, attitudes, beliefs, opinions) that are psychologically inconsistent with each other. Stated differently, two cognitions are dissonant if, considering these two cognitions alone, the opposite of one follows from the other....Because the experience of cognitive dissonance is unpleasant, people are motivated to reduce it either by changing one or both cognitions in such a way as to render them more compatible (more consonant) with each other, or by adding more cognitions that help bridge the gap between the original cognitions."

*traveling component. The vertically traveling component is then reflected from a totally reflecting mirrors so that both now move horizontally to the right. At a time when both components are surely within the region of the two boxes, the doors are closed and the two components of the original wavepacket reflect back and forth from the far and near ends of each box. The box pairs are positioned in front of a screen that will flash when slits are opened in the boxes as discussed below. A practical realization for photons is readily accomplished; for neutrons or atoms it would be more difficult but not technically out of the question. Long horizontal arms could substitute for boxes; actual containment is not needed for our argument[3].]* (6)

For the first set of box pairs, the Demonstrator instructs the Experimenter to determine which box of each pair holds the small object by opening a slit in first one box of a pair and then the other. He places each box pair in turn in front of a screen which will flash indicating the impact and presence of an object. (Later the Experimenter will examine the object in more detail.) About half the time he notes a flash at some approximately random place on the screen indicating that an object impacted there. Opening a slit in the second box of that pair, he sees no flash. On the other hand, if he sees no flash on opening the first box of a pair, there is always a flash on opening the second box of that pair.

The Experimentalist is admittedly not *completely* theory neutral. For one thing, he is not a solipsist. He holds to the logically unprovable thesis that a physically real world exists beyond his senses and that his senses inform him about that independently existing real world. Therefore tentatively identifying the detection of an object emerging from a box with its immediate prior existence there, the Experimenter tentatively concludes that for this set of box pairs, one box of each pair contained the object, and the other box of that pair was empty.

Presented with a second set of box pairs, the Experimenter, following the Demonstrator's different instructions, places a box pair in front of the screen and opens slits in both boxes at approximately the *same time*. There is a flash on the screen. Repeating the procedure with further box pairs in the same position, a flash appears somewhere on the screen for each box pair opened. But this time the Experimenter finds a pattern in the flashes: many flashes appear at some places on the screen, at other places there are none. Each object impacting from this second set of box pairs follows a rule allowing it to land only in a set of specific regions. In which particular region an object lands is random, but the statistics are not now our issue. The point is that in this case each and every individual object obeys the rule allowing it to land in certain places and forbidding it from landing in other places.

By opening sets of box pairs with different separations between the boxes of each pair, the Experimenter discovers the rule: the spacing of the screen regions where the objects impact is inversely proportional to the separation of the boxes. Each object thus obeys a rule depending on the *separation* of its box pair. Each object "knows" the separation of its box pair. The Experimenter thus concludes that for *this* set of box pairs some aspect of each of the objects must have been present in both boxes of its pair.

---

[3] This experimental arrangement is similar to that of Wheeler's "delayed-choice" experiment[6], although the discussion there is in terms of the quantum theory. The delayed-choice aspect could be noted here: the Experimenter's Heisenberg choice could be made after the object had moved beyond the half-transmitting mirror.

*[The Experimenter is apparently dealing with an interference phenomenon. However, we wish to be as theory neutral as possible, and the existence of a rule related to the box spacing, which is followed by each particle, is sufficient to make our point.]*

The Experimenter, now presented with further sets of box pairs, is told that for each set he may choose *either* of the two previous experiments. Whenever he chooses to open the box pairs sequentially, he appears to demonstrate that for this particular set of box pairs each object was wholly in a *single* box. Whenever he chooses to open the two boxes of a pair simultaneously, he appears to demonstrate that some aspect of each object was in *both* boxes of its pair. Or, in other words, the object somehow "knew" the separation of both boxes.

*[How simultaneous the openings need be would depend on details such as the slit-width. As the time between openings increases from zero, the pattern gradually washes out. But no matter, any rule dependent on box-pair spacing leads to the same conclusion.]*

The physical properties of the object would thus appear paradoxically determined by the Experimenter's subjectively free choice of what to observe. (These experimental outcomes are nevertheless objective in the sense that all observers see the same results.) It would seem that if the Experimenter had made the opposite choice, he would have established a prior condition for the objects inconsistent with the one actually established. He is puzzled.

The Experimenter's plausible guess is that some physical aspect of his simultaneous or sequential opening of the boxes of each pair produced what *appeared* to be contradictory prior conditions. Perhaps the physical action in opening the boxes caused the object to be in a single box or be distributed over both. Or perhaps the *apparently* empty box actually contains a physical entity which can guide the object's motion to produce the observed pattern. The Experimenter will investigate further.

*[Were the Experimenter versed in quantum theory, he would contemplate parts of a wavefunction associated with each box. Were he also of a Bohmian bent, he would envision a quantum potential in and around each box. But we think of the Experimenter as an extremely competent, open-minded observer unprejudiced by any particular theoretical doctrine.]*

*Further Investigations*: The Experimenter finds all opening techniques, slow ones, rapid ones, or extremely gentle ones produce the same results. Even techniques for finding which box contains the object which do not require opening (weighing or peeking, for example) produce the same results. No matter how the Experimenter obtains "which box" information, a whole object is found in a single box.

On painstaking examination, the objects on the screen appear identical in every respect to the objects initially sent into the box pairs, regardless of which experiment was done, even if only a single box of a pair was opened. To determine that the entity being examined is the *entire* object sent into the box pairs, the apparently empty box and the surrounding region is exhaustively searched for any field or other physical entity. Absolutely nothing is found.

An examination of a box which turned out to be empty should therefore not disturb the object--even though it did give "which box" information. This suggests the following test. The Experimenter opens one box of each box pair, and in the approximately half the cases in which the object was found to be in the box opened, he discards those box pairs. For the remaining box pairs, the opened box was reclosed. With this set of presumably physically *undisturbed* objects he performs the simultaneous opening experiment. No structured pattern appears. Again, it seems that merely the observer's obtaining the *knowledge* that the object was in the unopened box affected the situation. Only when such information was *not* obtained (nor associated with another entity from which such information could later be obtained) could the pattern indicating aspects of the object in both boxes be formed.

*[This is the closest the Experimenter gets to displaying the **non**-local aspect of observation (which is more explicitly seen in quantum experiments with multiple objects with correlated properties). Within the quantum theory, however, the collapse of the wavefunction entails non-locality even for the single object.]*

Repeating these observations with increasingly large objects, the Experimenter gets the same results. He finds that the alignment and measurement must be done with ever greater precision, but only technology and budget seem to prevent his moving further up in scale. He cannot establish that these paradoxical results are confined to tiny objects.

*[If quantum mechanics is completely correct, the Experimenter's investigations seeking a physical mechanism for bringing about the observed phenomena will always prove fruitless.]*

As a non-solipsist, the Experimenter had assumed that prior to his observation there was a reality to the physical world, and, in particular, a matter of fact to the condition of the object in the box pairs. In fact he was more specific: he had assumed that the object was in one of two conditions: either it was *wholly* within a single box *or* that aspects of it were distributed over *both* boxes, two apparently contradictory situations. However, since investigations accepted as exhaustive failed to reveal any physical mechanism for creating a particular one of the two inconsistent observed realities, he is moved to consider the possibility that the observed reality was somehow created by the observation *itself*, that the observed reality is created *solely* by the observer's acquisition of knowledge. If so, the observer is inseparably involved with the observed system. That would challenge his view of a physically real world existing independently of his senses perceiving it.

The only alternative the Experimenter sees to this observer-involved reality is to question his ability to freely choose the experiment. Thus if he assumes that no investigation will *ever* reveal a physical mechanism for bringing about one of the two inconsistent situations arising solely from his apparently free choice of experiment, he is faced with either accepting an observer involvement, or denying his free choice. *Neither* notion is acceptable to him. He thus experiences a cognitive dissonance. This is a measurement problem intimately involving the observer, one arising *directly* from the **quantum** experiment.

*[An infinity of realities in addition to that which the Experimenter had assumed are allowed by **logic**. In fact, quantum theory says that there was indeed **no** matter of fact about the prior* *position of the object: it was an unobservable superposition state. Were the Experimenter apprised of the quantum theory, he would likely see a conflict between the various possible observed realities given by the theory and the single observed actuality. He would thereby transfer his measurement problem from the experiment, where it first arose, to the theory. We discuss such a transfer in Section 5, where we discuss the measurement problem as it arises in the theory with the Dirac choice.]*

## 4.  BUT A *ROBOT* CAN DO IT!

A not unreasonable response to the argument that a measurement problem involving the observer arises in the quantum experiment is that a mechanical robot can do the experiment. It could then print out a report of its results. That report would be indistinguishable from one produced by the human Experimenter. The claim could then be that there was no involvement of an observer.

Since it is a human observer who must decide the nature of any measurement problem that exists, let's consider that robot-performed experiment from the point of view of the Experimenter. How would he interpret the robot's report? The robot's print-out merely indicates that box-pair sets number 1, 3, 8, 10, and 11 contained objects which were wholly in one box, while sets number 2, 4, 5, 6, 7, 9, and 12 contained objects which were distributed over both boxes. In itself, this print-out presents no puzzle. The Experimenter could assume that the respective box-pair sets indeed contained objects of just such kinds. But suppose he learns that the robot's "decision" of which experiment to do with each box-pair set is determined by a supposedly random coin flip. Heads, it opens the box pairs sequentially, tails simultaneously. Now something is puzzling: the coin's landing is correlated with what was in a particular box-pair set.

The Experimenter's exhaustive search for a physical mechanism correlating the coin flip with the contents of the box-pair set is fruitless. But to exclude any physical correlation at all, the Experimenter selects the robot's slit-opening procedure with the decision method he is most sure has no physical connection with the box pairs: his own supposedly free choice. He now experiences his original bafflement. The robot argument demonstrates that the role of the observer can be hidden in an unexplained phenomenon (in this case, the correlation of the coin flip with a box-pair result). From the Experimenter's point of view, the cognitive dissonance is still there, and a measurement problem arising in the experiment involves the observer.

*[One can consider the robot situation from the perspective of a quantum theorist. In this view, by opening the boxes in a particular way, the robot entangles the wavefunction of the object with the environment, and the resulting decoherence brings about the classical-like probability of the actually observed events--i.e., the object being in a particular box or in a particular interference maximum. No observer need be considered--for all* **practical** *purposes. (Although, it is generally admitted that the question of the* **ultimate** *observer still remains. See reference 15, for example.) This point of view transfers the measurement problem from the Heisenberg choice (the free choice by the Experimenter of the particular experiment) to the Dirac choice (the apparently random choice by Nature of which of the possibilities represented by the wavefunction becomes the single actuality seen by the observer). Our point here is that a*

*measurement problem involving the observer is seen in the theory-neutral experiment* **as well as** *being seen within the quantum theory, and the robot argument does not resolve it.]*

## 5. DISCUSSION

Realistic versions of the experiments described in Section 3 can readily be done with small objects. We can count single photons, and we can see individual atoms with light. We can even pick them up and put them down. Interference is being demonstrated with increasingly large objects, recently with $C_{60}$ buckyballs[7] and perhaps soon with Bose-Einstein condensates. As the technology advances, spatial superposition states with more complex larger objects will no doubt be reported. Superposition states containing many billions of electrons have already been seen[8], and long-lived quantum entanglement has been demonstrated with macroscopic objects containing $10^{12}$ atoms[9]. There is, in principle, no limit to the size of objects that can display enigmatic quantum phenomena. (There are of course limits unlikely to ever be exceeded for practical reasons. The situation is analogous to the limits on the speed of a rocketship. We conceive and talk of rocketship speeds approaching that of light while for practical reasons such speeds are unlikely ever to be attained.)

As the experimental objects become more macroscopic, the bafflement arising directly from experiment becomes more compelling. We can no longer relegate paradoxical quantum phenomena to objects so small as to be considered mere "models," qualitatively different from objective physical entities like chairs and cats. Physics is increasingly pressed to confront the issue of observation, or at least acknowledge it as a hint that we have not yet told the whole story. If quantum mechanics is completely correct, the notion that we make free choices and the notion that a physical world exists independent of those choices confront each other in the quantum *experiment* to produce a measurement problem intimately involving the observer that arises independently of the quantum theory.

It is impossible to establish that we actually do make free choices. "Free will" may be an illusion, though it is almost undeniable. (I. B. Singer: "You have to believe in free will, you have no choice.") But the rejection of free choice needed to account for the experimental results is *not* simply that the experimenter's choices are determined by heredity and past experiences. The not-free choice here must include the remarkable correlation of observer's choice with physical phenomena, in the present example, with the object in the box pairs.

It is also impossible to establish there to be a physical reality prior to an observer's experiencing of it. It is even logically impossible to refute the extreme solipsistic position that there is no reality at all, that only sense impressions exist. But, rejecting anything like solipsism, the observation of an object in a particular place would seem to define what we mean by its existence there immediately prior to its observation. To the extent that the prior existence of objects is a meaningful concept at all, how could it more surely be established? (We are, of course, assuming our observations to be "gentle" in that they cannot physically create or destroy the object. A gamma ray causing pair creation is, for example, a not-gentle "observation" of an electron.)

We can try to be a bit more explicit about the Experimenter's view of reality. Our Experimenter, like most experimenters everywhere, had assumed that his observation reveals an immediately preexisting physical situation not brought about by his free choice of experiment. He later assumes that his fruitless search to detect a physical mechanism for the experimental results challenging that reality assumption was exhaustive. Moreover, he assumes that no physical mechanism which is *in principle undetectable* is a meaningful explanation.

This is an admittedly limited characterization of our Experimenter's view of reality. Were we proposing a *resolution* of the quantum measurement problem, a more complete discussion of "reality" would be in order. However, we merely describe the basis of our Experimenter's bafflement brought about by his experiment and his further investigations, which cause his beliefs in a prior reality and in free choice to conflict. His resulting bafflement presents him with a measurement problem. This bafflement is a measurement problem arising directly from the experiment and one which intimately involves the observer in a manner beyond anything seen in classical physics.

We note that a rejection of the above-described physical reality *prior* to its observation together with the absence of any physical mechanism bringing about the observed actuality implies the creation of that actuality *by* its observation. Since with a different Heisenberg choice of experiment by the Experimenter there would have been a different actuality, there is here an intrinsic role for the observer. (Some discussions of quantum mechanics consider it not meaningful to talk of what *would* have happened in experiments that were not in fact done. Denying the significance, or perhaps the possibility, of an alternative experiment comes close to denying free choice.)

The measurement problem arising from the quantum experiment does not necessarily imply that something "from the mind of the observer" affects the external physical world. The measurement problem does, however, hint that there is more to say about the physical world than quantum theory says. Any extension of the theory which predicts different results for the Experimenter's investigations would of course violate our assumption that his investigations could find nothing other than quantum theory's predictions.

Extensions such as those of Penrose or Stapp, mentioned above, explicitly include observer involvement. On the other hand, a recent report[10] argues that local hidden variables are not in fact excluded by Bell's theorem and the related experiments. An experimentally demonstrated existence of local hidden variables would not only violate our assumptions but could conceivably provide a solution to the measurement problem that removes observer involvement. Other than some extension of the quantum theory, we can envision only one alternative for addressing the measurement problem. We refer to it as "the psychological interpretation of quantum mechanics" and discuss it in Section 7.

*The measurement problems arising with the Heisenberg and Dirac choices*: In a quantum experiment, after the Experimenter makes the Heisenberg choice of which property of a system to observe or measure, Nature makes the Dirac choice of the particular observed outcome or measured value. The Dirac choice, that the object was, for example, found in a *particular* box, or that it landed in a *particular* maximum, need not seem paradoxical to the Experimenter. He could just assume that the objects were incompletely characterized. Baffled by the measurement problem posed by his Heisenberg choices, the theory-neutral Experimenter might well not focus on a problem posed by Nature's Dirac choice.

On the other hand, a quantum theorist, knowing that the linearity of the Schrödinger equation allows no mechanism for the theory to select a particular actuality from the multiplicity of superposed possibilities, would likely see things differently. The theorist may consider the superposition to be the physical reality existing prior to an observation. The measurement problem, the bafflement, here arises from the conflict between the theorist's belief that the theory's set of possibilities represents the physical world and the fact that what one actually observes is so different.

The quantum theorist might well not focus on the measurement problem arising from the Heisenberg choice of experiment. Though that choice determines the basis in which the state is expanded and thus the set of particular outcomes, since all bases are mathematically equivalent, this choice might well not be the focus of his attention.

Allow us to further contrast the measurement problems due to the Dirac and Heisenberg choices with a bit of fantasy inspired by a parable due to Omnès[11]. On the higher plane on which they dwell, Everettians experience many worlds, the multitude of simultaneous realities given by quantum theory. On their higher plane, this is readily seen as the nature of physical reality, and they thus experience no measurement problem. One young Everettian, sent down to explore planet Earth, was shocked to find his simultaneous multiple realities collapse to a single sharp actuality. Curiosity impelled repeated descents. On each return to Earth, he randomly perceived a collapse to a single one of the many realities he was accustomed to experience simultaneously on his higher plane. Baffled by an observation unexplainable by the quantum physics he understood so well, he reported there to be a measurement problem on the lower Earthly plane in Nature's selection of a single actuality (a Dirac choice) .

Our Everettian had a favorite way of viewing the multiple realities he experienced. He understood, however, that this personal choice, his point of view, or his basis, was mathematically equivalent to an infinite number of other possible bases. In a rather unusual mood on a particular descent to Earth, our Everettian adopted a different basis for viewing reality. He experienced a *second* shock. With his new basis, the collapse was not only to a single actuality, something he had by now gotten quite accustomed to, but the collapse was to an actuality that was logically inconsistent with a previously experienced actuality. He had to report a *new* measurement problem arising from his choice of basis (a Heisenberg choice).

*Restricting the domain of the theory:* **Could the intrusion of the observer be resolved by restricting the domain of the theory as was done in classical physics?** Classical physics resolved its observer problem, the logical conflict between determinism and free choice, by excluding the freely-choosing mind of the observer from the domain treated by physical theory. That was possible with consistency because the picture of reality given by one possible classical experiment never conflicted logically with that of a different experiment that might have been chosen. The problematic issue of the observer's free choice arose only within the classical *theory*, with determinism; it never arose in classical *experimental* observations--as it does in quantum experiments.

The domain of quantum theory can in fact also be restricted to exclude only the mind of the observer. That is essentially what von Neumann did[12]. But according to such analysis, the world encompassed by the quantum theory still includes all the possibilities prior to the Dirac choice, and only by an unspecified process beyond the theory, explicitly involving the mind of the observer, does the wavefunction's many possibilities collapse to the single observed result. The observer thus still intrudes into the experiment, if not into the theory.

## 6. COMMENTS ON INTERPRETATIONS OF QUANTUM MECHANICS

The confounding nature of observation continues to stimulate contending interpretations of quantum mechanics. Almost all interpretations address the aspect of the measurement problem arising within the theory, from the conflict between the many possibilities represented by the wavefunction and the single observed actuality resulting from Nature's Dirac choice. Since our focus is not there, but rather on the measurement problem arising from the *experiment*--independently of the quantum theory--our brief discussion of these interpretations is an aside to the main point of this paper. We do not intend a comprehensive review of these interpretations. With thumbnail sketches of several current interpretations, we merely indicate how each involves those beliefs of our Experimenter that conflict to present a Heisenberg-choice measurement problem involving the observer.

*Copenhagen*: According to the pragmatic version of the Copenhagen interpretation implicit in most quantum mechanics texts, all an Experimenter ever need know are the outcomes of macroscopic observations made with essentially classical equipment. The Experimenter therefore need consider no microscopic reality prior to its classical observation. Such "reality" can be assumed non-existent, or at least irrelevant. The observer's conflict between reality and free choice thus evaporates. Since the observer is classical, the issue of free choice can logically be ignored, as in classical physics. This version of the Copenhagen interpretation presents us with the fundamentally unsatisfying situation of one physics for the microscopic world and a different physics for the macroscopic world. This divided universe is increasingly seen as an unresolved problem as experiments increasingly probe the ill-defined boundary between these two worlds. The von Neumann version of the Copenhagen interpretation divides the universe only at the observer and thus explicitly involves the observer in collapsing the wavefunction. A recent interpretation which might be considered a version of the Copenhagen interpretation abandons the notion of particles (or fields) entirely to treat nothing but actual observations[13].
.

*Decoherence*: The decoherence approach discussed by Zurek[14] and others emphasizes that a macroscopic system is never isolated from the environment. The off-diagonal elements of the density matrix of the microscopic quantum system coupled with the measuring apparatus as the measuring apparatus interacts with the environment in a measurement process are shown to rapidly average out, resulting in the density matrix rapidly coming to appear identical to a classical probability formula.

Before the measurement interaction, this approach accepts the same lack of prior reality as the Copenhagen interpretation. Moreover, in principle, though the density matrix develops the form of a classical probability, it is not one. In a later paper[15] Zurek does refer to the conscious observer's involvement in the *ultimate* "collapse."

*Consistent Histories*: The consistent histories (or decoherent histories) interpretation[16] holds quantum mechanics to be a fundamentally stochastic or probabilistic theory. Decoherence generalizes and replaces the notion of "measurement" and independence of the observer is obtained by assuming the existence of a quasi-classical world. Considering the existence of two inconsistent histories (e.g., considering both our which-box experiment and our interference experiment for the same object ) is "meaningless." The actuality finally observed defines the previously existing situation. In this interpretation there thus appears to be no reality of the kind our experimenter holds to.

*Many Worlds*: In the "many worlds" interpretation, originally formulated by Everett[17], all physically possible results actually obtain ("many worlds")--or at least exist in the mind of an experimenter ("many minds")[18]. The experimenter chooses to do all possible experiments and experiences all possible results. Supposedly no new physics is introduced. It is, however, unclear how probabilities for different possibilities arise, precisely what constitutes an observation, or what determines the preferred basis for observations. There is no free choice in the usual sense since all possible choices are made. There is also no *particular* prior reality--the multitude of prior realities implied by the wavefunction are all equally actual, each defining its own world, each world including a version of the experimenter.

*Ithaca*: A recent interpretation of quantum mechanics, proposed by Mermin and called the Ithaca interpretation[19], modestly addresses only a part of the measurement problem. In this interpretation there is "reality," and then there is "*physical* reality." Reality includes the perception of events. Physical reality, that aspect of reality described by quantum theory, includes only correlations. The things correlated do not have physical reality and need not be the proper study of physics. This approach explicitly does not address, nor does it deny, the involvement of the observer.

*Spontaneous Localization*: Two approaches to the observer problem explicitly add new physics to quantum mechanics, new physics that is in-principle testable. They thus violate a basic assumption of our argument, and are not, strictly speaking, merely interpretations. Ghirardi, Rimini, and Weber[20] propose nonlinear, nonlocal modifications of the Schrödinger equation that would rapidly collapse spatially distinct superposition states to a single actuality for macroscopic objects. If the object in our box pairs were large, it would collapse into a single box before any interference experiment could be done. This interesting speculation removes observer involvement for all *practical* purposes. For observations of microscopic objects, at times shorter than the spontaneous localization time, the observer is, in principle at least, still involved. A different localization approach due to Penrose[21] explicitly involves the observer. Here gravitational interaction in the brain of the observer causes superpositions to collapse to actuality, and this collapse is also responsible for a conscious experience.

*Stapp*: Unlike interpretations seeking to minimize the intrusion of the observer into physics, Stapp emphasizes the role of the observer[22]. Stapp extends the von Neumann version of the Copenhagen interpretation to provide a comprehensive quantum theory of consciousness. The state vector of the universe is the representation of an objectively existing information structure which evolves according to the Schrödinger equation and comprises physical reality. The mind of an observer (a primitive of the theory) makes a Heisenberg choice by posing a particular question, thus choosing a basis for an observation. Nature then makes a probabilistic Dirac choice collapsing the state vector of the universe.

*Quantum Potential*: Defying Copenhagen's pragmatic stance, Bohm presented an ontological, and completely deterministic, interpretation[23]. Although its complete relativistic extension is still problematic[24], with one natural-seeming assumption all statistical results can agree with those of Schrödinger-equation quantum theory, though they need not[25]. To the extent that this version of quantum mechanics predicts new physical phenomena, it evades one of our assumptions leading to observer involvement. But let us assume that the quantum potential program can be brought to fruition and *completely* duplicates the predictions of present quantum theory.

In this interpretation, the wavefunction is not a complete description of reality; actual particles exist and have actual positions. There is, for example, a real object in one or the other of our boxes. In addition, a "quantum potential," derived from the Schrödinger equation, exists in the space in and around both boxes and provides a force to guide each object to a predetermined position in an interference experiment. In a multiparticle system, the quantum potential is a function of the coordinates of all the particles, and a change in the situation of any particle instantaneously influences all others, however remote. This instantaneous binding brings about the nonlocal effect of observation (which arises with wavefunction collapse, involving the observer in some other interpretations).

The Bohm interpretation postulates a potential detectable *only* by its bringing about the results predicted by the Schrödinger equation. Being derived from that equation, it is, of course, mathematically constrained to do just that. Such a universally pervasive and instantaneously physically efficacious field certainly conflicts with our Experimenter's view of reality. But would an entity with no other role than bringing about the results of Schrödinger quantum mechanics resolve the measurement problem arising from his apparently paradoxical experimental observations?

As analogies, if the neutrino had *no* role other than its original one of preserving energy and momentum conservation, or if the speed of light being the same in all inertial frames had *no* testable consequence other than "explaining" the null result of Michelson-Morley type experiments, these postulates would hardly be accepted as resolutions of experimental paradoxes.

Closer to the present situation, if the A-vector had no other consequence beyond the results predicted by Maxwell's equations, it would be regarded as a calculational device and not a physical entity. Since there can be an A vector in regions in which there is no electric or magnetic field, within Maxwell theory its presence there is completely undetectable. The physical reality of the A vector was established only by the detection of the Bohm-Aharanov effect, a prediction of a quantum extension of Maxwell theory. The quantum potential can be considered analogous to the A-vector before the Bohm-Aharanov effect: if it predicts no phenomena beyond Schrodinger quantum mechanics, it could be considered merely a calculational device. It would not resolve the Experimenter's measurement problem.

Nevertheless, as was initially the case for the postulates of the neutrino, the constancy of the speed of light, and the A-vector, the quantum potential is suggestive of new physical phenomena relating the microscopic quantum world to the macroscopic world and to, perhaps, the observer. In his later work, in which substantial new physical phenomena are suggested, Bohm goes beyond the divided universe of the Copenhagen interpretation to discuss the observer in an *un*divided universe[26].

We finally re-emphasize that our brief sketches of how various interpretations relate to the assumptions of the Experimenter is an aside to our main argument: i.e., since the involvement of the observer arises directly from theory-neutral experimental observations logically prior to the theory, no interpretation of the quantum *theory* can resolve the measurement problem by removing the observer, as is possible in classical physics. The number of interpretations in current contention addressing the measurement problem emphasizes that a resolution still eludes us.

## 7. CONCLUSIONS

Though quantum mechanics is a fully consistent theory and sufficient as a useful guide to the physical phenomena around us, we may wish more than an algorithm for computing probabilities. Classical physics provided more; it imparted a worldview, but one we now know to be fundamentally flawed.

The observations encompassed by classical physics allowed the exclusion of the observer from the universe addressed by physics. The worldview suggested by the quantum *experiment* either challenges that exclusion or suggests new physical phenomena. It not only hints at a different view of reality, but "[It is] likely that the new way of seeing things will involve an imaginative leap that will astonish us."[27]

Physicists appropriately seek the least astonishing solution. However, recent comments on the observer problem can give the impression that the issue has been resolved, that no hints of a deeper mystery are present. It is a temptation (that we can share) to reject observer involvement as being so preposterous that no conceivable evidence could ever establish it. But such a stance hardly seems open-minded, and the history of science suggests it is flawed.

Since the observer problem in classical physics (the conflict of free will with determinism) arose only *within* the theory, the problem could be evaded by excluding consideration of the mind of the observer from the realm encompassed *by* the theory. This option seems unavailable to quantum physics since the intrusion of the observer occurs in the experimental observations. To avoid the observer in this case, the excluded realm must be greater and its boundary more vague. If, indeed, the reality that physics addresses is only part of a larger interacting reality including the observer, quantum physics experiments have disclosed physical evidence for the existence of such a larger reality.

Is this reaching too far? The measurement problem seen in the quantum experiment is the bafflement arising from the conflict between the belief in free choice and the belief in the existence of a physical reality prior to that choice. If that measurement problem has no consequence other than such bafflement, is it nothing but a psychological problem? In a sense it is. So, of course, is the more frequently addressed version of the measurement problem arising within the quantum theory. In this latter case, the belief in the uniqueness of discrete events (denied, for example, by the many worlds interpretation) conflicts with our belief (as physicists, at least) that the mathematical structure of a theory corresponds to a physical reality. Both versions of the measurement problem can be seen as arising from a cognitive dissonance.

One may therefore argue that no *physical* measurement problem actually exists; Nature need not correspond to even the most basic human intuitions. This attitude is probably close to that of the majority of physicists. Accepting such a position, our problem with observation becomes an issue in psychology worth serious investigation. Namely, what is it about the human mind that Nature's most fundamental law, quantum mechanics, creates such cognitive dissonance? An answer could be called the "psychological interpretation of quantum mechanics." That we evolved in a world with classical physics being a good approximation is not a sufficient explanation. We evolved in a world where things moved slowly compared to the speed of light. But we find no need for "interpretations" of relativity. The bafflement with quantum mechanics is unique. However, since this way around the quantum measurement problem is applicable to *any* paradoxical situation, it warrants a skeptical view.

The quantum measurement problem presents a worldview profoundly different from that suggested by classical physics. That the intrusion of the observer occurs at the level of the experiment, and arises independently of any interpretation of the theory, emphasizes this conclusion.

At this stage, where no testable modification of the quantum theory convincingly resolves the enigmatic role of the observer, and where "a model entity whose states correspond to a reasonable caricature of conscious awareness...[is] well beyond what is conventionally considered physics,"[28] the intrusion of the observer into the physical world is not readily confronted with the techniques familiar to physicists. For all *practical* purposes no confrontation seems required. However, physics has conceivably encountered a phenomenon whose significance should not be minimized.

**ACKNOWLEDGMENTS**

We thank Donald Coyne, Charles Crummer, Gaston Fischer, Stanley Klein, Alex Moraru, and Henry Stapp for helpful comments on drafts of this paper.